\documentclass[english,preprint,showpacs,preprintnumbers]{revtex4}
\usepackage[latin9]{inputenc}
\usepackage{amsmath}
\usepackage{amssymb}

\makeatletter

\providecommand{\tabularnewline}{\\}

\@ifundefined{textcolor}{}
{%
 \definecolor{BLACK}{gray}{0}
 \definecolor{WHITE}{gray}{1}
 \definecolor{RED}{rgb}{1,0,0}
 \definecolor{GREEN}{rgb}{0,1,0}
 \definecolor{BLUE}{rgb}{0,0,1}
 \definecolor{CYAN}{cmyk}{1,0,0,0}
 \definecolor{MAGENTA}{cmyk}{0,1,0,0}
 \definecolor{YELLOW}{cmyk}{0,0,1,0}
}

\usepackage{babel}

\usepackage{amsfonts}
\usepackage{graphics}
\usepackage{epsfig}
\usepackage{bm}

\makeatother

\usepackage{babel}
\begin{document}

\title{Accidental symmetries and massless quarks in the economical 3-3-1
model}

\author{J. C. Montero}

\email{montero@ift.unesp.br}

\selectlanguage{english}%

\affiliation{Instituto de Física Teórica, Universidade Estadual Paulista, R. Dr.
Bento Teobaldo Ferraz 271, Barra Funda, SP, 01140-070, Brazil. }

\author{B. L. Sánchez--Vega}

\email{brucesanchez@anl.gov}

\selectlanguage{english}%

\affiliation{Argonne National Laboratory, 9700 S. Cass Avenue, Argonne, IL 60439. }
\begin{abstract}
In the framework of a 3-3-1 model with a minimal scalar sector, known
as the economical 3-3-1 model, we study its capabilities of generating
realistic quark masses. After a detailed study of the symmetries of
the model, before and after the spontaneous symmetry breaking, we
find a remaining axial symmetry that prevents some quarks to gain
mass at all orders in perturbation theory. Since this accidental symmetry
is anomalous, we also consider briefly the possibility to generate
their masses for non-perturbative effects. However, we find that non-perturbative
effects are not enough to generate the measured masses for that three
massless quarks. Hence, these results imply that the economical 3-3-1
model is not a realistic description of the electroweak interaction
and it has to be modified. 
\end{abstract}
\maketitle

\section{Introduction}

Up to now, from the experimental point of view, neutrino masses and
their mixing, and dark matter are the only issues demanding for explanations
beyond the standard model (SM). On the other hand, from the theoretical
point of view, the quest for a more deeper understanding lead us to
believe that a more fundamental model of the interactions is needed.
That model should be able to answer simple but deep questions. Some
of these questions are: why the number of families of quarks and leptons
is three? Is there a more fundamental relation (symmetry) between
quarks and leptons? Why does the observed pattern for the particle
masses have this particular form? Should not the parameters involved
have any calculability? What is the origin of CP violation? Even in
the SM, what is the origin of the CP violating CKM phase? Can it be
computed? Is there any more efficient mechanism able to account for
the matter-antimatter asymmetry in the Universe? What is the mechanism
that generates masses and mixing angles for neutrinos? Is there CP
violation in leptons? What would be its role in the evolution of the
Universe? How dark matter and dark energy can be incorporated? Unfortunately,
up to now, the experimental efforts were not able to indicate exactly
what the physics beyond the SM should be.

In the framework of gauge theories, one way of introducing new physics
is to consider a gauge symmetry group larger than the SM one. Some
years ago models with the $\ensuremath{\textrm{SU}(3)_{C}\otimes\textrm{SU}(3)_{L}\otimes\textrm{U}(1)_{X}}$
gauge symmetry were proposed \cite{Pisano:1991ee,Frampton:1992wt,Foot:1992rh,PhysRevD.47.2918},
which are having considerable further developments. The so-called
3-3-1 models present interesting features concerning the asked questions
above. One of them is that, depending on the representation content,
the triangle anomalies cancel out, and the number of families has
to be a multiple of three. More precisely, it must be just three due
to the asymptotic freedom. A version of this kind of models called
minimal \citep{Pisano:1991ee,Frampton:1992wt,Foot:1992rh,PhysRevD.47.2918}
presents a Landau-like pole when $\ensuremath{\sin^{2}\theta_{W}=1/4}$
at energies of the order of a few TeVs \citep{Dias:2005}. This particular
behavior stabilizes the electroweak scale avoiding the hierarchy problem
and also explains why it is observed $\ensuremath{\sin^{2}\theta_{W}<1/4}$.
This model also accounts for the electric charge quantization independently
of the nature of the massive neutrinos, i.e. whether they are Dirac
or Majorana particles \citep{Pires:1998}. The model has also interesting
features concerning the strong CP problem. In the minimal 3-3-1 model
there is an almost automatic Peccei-Quinn (PQ) symmetry, and an automatic
one in the so called economical version of the model, as we will show
here below. In both versions there are ways of solving the strong
CP problem while keeping the corresponding axion invisible and protected
against gravitational effects \citep{Dias:2004,Bruce:2011PQ}. Due
to a larger gauge symmetry group and a rich scalar sector, this kind
of model has called some attention in many other subjects like new
sources of CP violation, active neutrino mass generation and mixing,
dark matter candidates and $\ensuremath{Z^{\prime}}$-boson physics.

In this paper we are concerned with the quark mass generation, in
the context of the economical 3-3-1 model (E331, for short). In particular
we investigate the capabilities of the model in generating realistic
quark masses. The quark sector of this model was already considered
in literature and conflicting results were found \citep{Bruce:2011PQ,Dong:2012bf}.
In this work, in order to clarify this important issue, we do a detailed
study of the symmetries (local and global symmetries) of the entire
E331-model Lagrangian. Once we have identified all the symmetries,
and after the spontaneous symmetry breaking (SSB) of the scalar potential,
we investigate which are the remaining symmetries (if any) of the
vacuum state. In other words, we seek which are (if any) the independent
linear combinations of the group generators that annihilate the vacuum
state, in order to know if the corresponding symmetries are realized
\emph{à la} Wigner-Weyl (WW) or Nambu-Goldstone (NG). That is of fundamental
importance since it will affect the physical particle spectrum. If
the total Lagrangian and vacuum state are both invariant under a symmetry
transformation, this is a WW realization of that symmetry. On the
other hand, if the vacuum is not invariant this is a NG realization
and this implies a massless NG scalar boson. We find that there is
a WW realization of a subgroup of the initial symmetry group that
protects some quarks from getting mass at all orders in perturbation
theory, as expected from quantum field theory. 

The paper is organized as follows. In Sec. II we briefly review the
economical 3-3-1 model. In Sec. III we make a detailed study of the
symmetries of the model, in both situations before and after the spontaneous
symmetry breakdown, and its implication for the quark masses. Non-perturbative
effects contributing to quark masses are also briefly considered.
Our conclusions are presented in Sec. IV.

\section{A brief review of the economical 3-3-1 model}

The model considered has a matter content given by \citep{Ponce:2002sg}
\begin{align}
\Psi_{aL} & =\left(\nu_{a},e_{a},\left(\nu_{aR}\right)^{C}\right)_{L}^{T}\sim\left(\mathbf{1},\mathbf{3},-1/3\right)\text{,}\ e_{aR}\sim\left(\mathbf{1},\mathbf{1},-1\right)\text{,}\notag\nonumber \\
Q_{\alpha L} & =\left(d_{\alpha},u_{\alpha},d_{\alpha}^{\prime}\right)_{L}^{T}\sim\left(\mathbf{3},\mathbf{3}^{\ast},0\right)\text{,}\ Q_{3L}=\left(u_{3},d_{3},u_{3}^{\prime}\right)_{L}^{T}\sim\left(\mathbf{3},\mathbf{3},1/3\right)\text{,}\notag\nonumber \\
u_{aR} & \sim\left(\mathbf{3},\mathbf{1},2/3\right)\text{,}\ u_{3R}^{\prime}\sim\left(\mathbf{3},\mathbf{1},2/3\right),\ d_{aR}\sim(\mathbf{3},\mathbf{1},-1/3)\text{,}\ d_{\alpha R}^{\prime}\sim\left(\mathbf{3},\mathbf{1},-1/3\right),\nonumber \\
\chi & =\left(\chi^{0},\chi^{-},\chi_{1}^{0}\right)^{T}\sim\left(\mathbf{1},\mathbf{3},-1/3\right),\ \rho=\left(\rho^{+},\text{ }\rho^{0},\text{ }\rho_{1}^{+}\right)^{T}\sim\left(\mathbf{1},\mathbf{3},2/3\right),\label{matter}
\end{align}
where $a=1$, $2$, $3$, $\alpha=1$, $2$ and the values in parentheses
denote respectively the quantum numbers corresponding to the $\left(\textrm{SU(3)}_{C},\,\textrm{SU(3)}_{L},\,\textrm{U(1)}_{X}\right)$
groups. From now on Latin and Greek letters always take the values
$1$, $2$, $3$ and $1$, $2$, respectively. 

With the quark, lepton and scalar multiplets in Eq. (\ref{matter})
we have that the most general Yukawa interactions allowed by the gauge
symmetries and renormalizability are 
\begin{align}
\mathcal{L}_{\text{Y}} & =Y_{ab}\overline{\Psi_{aL}}e_{bR}\rho+Y_{ab}^{\prime}\epsilon^{ijk}\left(\overline{\Psi_{aL}}\right)_{i}\left(\Psi_{bL}\right)_{j}^{C}\left(\rho^{\ast}\right)_{k}\nonumber \\
 & +G^{1}\overline{Q_{3L}}u_{3R}^{\prime}\chi+G_{\alpha\beta}^{2}\overline{Q_{\alpha L}}d_{\beta R}^{\prime}\chi^{\ast}+G_{a}^{3}\overline{Q_{3L}}d_{aR}\rho+G_{\alpha a}^{4}\overline{Q_{\alpha L}}u_{aR}\rho^{\ast}\nonumber \\
 & +G_{a}^{5}\overline{Q_{3L}}u_{aR}\chi+G_{\alpha a}^{6}\overline{Q_{\alpha L}}d_{aR}\chi^{\ast}+G_{\alpha}^{7}\overline{Q_{3L}}d_{\alpha R}^{\prime}\rho+G_{\alpha}^{8}\overline{Q_{\alpha L}}u_{3R}^{\prime}\rho^{\ast}+\text{h.c.,\label{lagyuk}}
\end{align}
where $G^{i}$, $Y_{ab}$ and $Y_{ab}^{\prime}$ are arbitrary complex
matrices and $Y_{ab}^{\prime}$ is also antisymmetric. We use the
convention that addition over repeated indices is implied. 

The $\chi$ and $\rho$ scalar multiplets break down spontaneously
the $\textrm{SU(3)}_{C}\otimes\textrm{SU(3)}_{L}\otimes\textrm{U(1)}_{X}$
gauge symmetry. The vacuum expectation values, VEVs, in this model
satisfy $\left\langle \text{Re\thinspace}\rho^{0}\right\rangle \equiv v\text{, }\left\langle \text{Re\thinspace}\chi^{0}\right\rangle \equiv u\ll\left\langle \text{Re\thinspace}\chi_{1}^{0}\right\rangle \equiv w\text{.}$
The most general scalar potential, that is both invariant under the
gauge symmetry and renormalizable, is 
\begin{align}
V & =\mu_{\chi}^{2}\chi^{\dagger}\chi+\mu_{\rho}^{2}\rho^{\dagger}\rho+\lambda_{1}\left(\chi^{\dagger}\chi\right)^{2}+\lambda_{2}\left(\rho^{\dagger}\rho\right)^{2}+\lambda_{3}\left(\chi^{\dagger}\chi\right)\left(\rho^{\dagger}\rho\right)+\lambda_{4}\left(\chi^{\dagger}\rho\right)\left(\rho^{\dagger}\chi\right).\label{potencial model a}
\end{align}
With only two scalar multiplets the scalar sector is simple and it
is, in principle, an appealing feature of this model comparing to
other 3-3-1 models \citep{Pisano:1991ee,Frampton:1992wt,Foot:1994ym}. 

Finally, the electric charge operator is written as 
\begin{equation}
Q=T_{3}-\frac{1}{\sqrt{3}}T_{8}+X\text{,}\label{chargeoperator}
\end{equation}
where $T_{3}$ and $T_{8}$ are the diagonal generators of the $\textrm{SU(3)}_{L}$
group and $X$ refers to the quantum number of the $\textrm{U(1)}_{X}$
group.

\section{Spontaneous symmetry breaking and massless quarks }

Before considering which symmetries are broken down, we look for all
exact symmetries, local and global, this model actually has. Doing
so, we realize that apart from the local gauge symmetry $\textrm{SU(3)}_{C}\otimes\textrm{SU(3)}_{L}\otimes\textrm{U(1)}_{X}$,
this model has two extra global $\textrm{U(1)}$ symmetries which
we denoted generically by $\textrm{U(1)}_{\zeta}$. In order to see
that, we write down the relations that these symmetries have to obey
in order to keep the entire Lagrangian invariant. From Eq. (\ref{lagyuk})
we obtain the following relations
\begin{eqnarray}
-\zeta_{Q_{3}}+\zeta_{u_{3R}^{\prime}}+\zeta_{\chi}=0, & -\zeta_{Q}+\zeta_{d_{R}^{\prime}}-\zeta_{\chi}=0\text{,} & -\zeta_{Q_{3}}+\zeta_{u_{R}}+\zeta_{\chi}=0,\label{e1}\\
-\zeta_{Q}+\zeta_{d_{R}}-\zeta_{\chi}=0, & -\zeta_{Q_{3}}+\zeta_{d_{R}}+\zeta_{\rho}=0, & -\zeta_{Q}+\zeta_{u_{R}}-\zeta_{\rho}=0,\\
-\zeta_{Q_{3}}+\zeta_{d_{R}^{\prime}}+\zeta_{\rho}=0, & -\zeta_{Q}+\zeta_{u_{3R}^{\prime}}-\zeta_{\rho}=0\text{,} & -\zeta_{\Psi}+\zeta_{e_{R}}+\zeta_{\rho}=0,\\
 & 2\zeta_{\Psi}+\zeta_{\rho}=0,\label{e5}
\end{eqnarray}
where the $\zeta_{\psi_{i}}$'s above denote the $\textrm{U(1)}_{\zeta}$
charges of the $\psi_{i}$ fields. Solving Eqs.~(\ref{e1}-\ref{e5}),
we find that all charges, $\zeta_{\psi_{i}}$, can be written in terms
of three independent ones. It means that the model has only three
independent $\textrm{U(1)}_{\zeta}$ symmetries. In principle, we
can choose whatever three independent $\textrm{U(1)}_{\zeta}$ symmetries
as basis. However, some physical considerations can be done to appropriately
choose them. First, we note that one of these symmetries is the $\textrm{U(1)}_{X}$
gauge symmetry which is anomaly free by construction and which has
an associated gauge boson. The other two are global symmetries and
they can be divided into a vectorial and a axial symmetry acting on
the quarks. The vectorial one is the well known baryon number symmetry,
denoted here as $\textrm{U(1)}_{B}$, which is an accidental symmetry
in this model as it is in the SM. The other one is a axial symmetry
also acting on the quarks, which we denote as $\textrm{U(1)}_{\textrm{PQ}}$.
The last symmetry is a PQ one since it is anomalous and $A_{\text{PQ}}$,
the coefficient of the $\left[\textrm{SU(3)}_{C}\right]^{2}\textrm{U(1)}_{\textrm{PQ}}$
anomaly, is $\varpropto-3\text{.}$ Also, notice that the $\textrm{U(1)}_{\textrm{PQ}}$
is a natural symmetry in the sense that it is not imposed, it follows
from the gauge local symmetry and renormalizability, instead. In other
words, the economical model naturally has a PQ symmetry. The assignment
of the three independent U$\left(1\right)$ quantum charges is shown
in Table \ref{U1charges} (these quantum charges appeared for the
first time in Ref. \citep{Bruce:2011PQ}, we have written them here
for the sake of completeness and clearness). Thus, the model actually
has a larger symmetry: $G\equiv\textrm{SU(3)}_{C}\otimes\textrm{SU(3)}_{L}\otimes\textrm{U(1)}_{X}\otimes\textrm{U(1)}_{B}\otimes\textrm{U(1)}_{\textrm{PQ}}$,
where the last two ones are accidental and global symmetries. 
\begin{table}[th]
\centering %
\begin{tabular}{|c|c|c|c|c|c|c|c|c|}
\hline 
 & \, $Q_{\alpha L}$  & \, $Q_{3L}$  & \, ($u_{aR}$, $u_{3R}^{\prime}$)  & \,($d_{aR}$, $d_{\alpha R}^{\prime}$)  & \, $\Psi_{aL}$  & \, $e_{aR}$  & \, $\rho$  & \, $\chi$ \tabularnewline
\hline 
\hline 
$\textrm{U\ensuremath{\left(1\right)}}_{X}$  & $0$  & $1/3$  & $2/3$  & $-1/3$  & $-1/3$  & $-1$  & $2/3$  & $-1/3$ \tabularnewline
\hline 
$\textrm{U\ensuremath{\left(1\right)}}_{B}$ & $1/3$  & $1/3$  & $1/3$  & $1/3$  & $0$  & $0$  & $0$  & $0$ \tabularnewline
\hline 
$\textrm{U\ensuremath{\left(1\right)}}_{\textrm{PQ}}$  & $-1$  & $1$  & $0$  & $0$  & $-1/2$  & $-3/2$  & $1$  & $1$ \tabularnewline
\hline 
\end{tabular}

\protect\caption{Assignment of the three independent U$\left(1\right)$ quantum charges
in the economical 3-3-1 model. \label{U1charges}}
\end{table}

Now, let us search for the remaining symmetries after the $\chi$
and $\rho$ scalar triplets obtain their VEVs, $\left\langle \chi\right\rangle \equiv V_{\chi}=\frac{1}{\sqrt{2}}\left(u,\,0,\, w\right)^{T}$
and $\left\langle \rho\right\rangle \equiv V_{\rho}=\frac{1}{\sqrt{2}}\left(0,\, v,\,0\right)^{T}$.
To do that, we consider an infinitesimal transformation of the total
group $G$ on the vacuum states to find the generators of the unbroken
subgroups as a linear combination of the $T_{i}$, $X$, PQ, and $B$
generators. Here, it is important to note that the $\textrm{SU(3)}_{C}\otimes\textrm{U(1)}_{B}$
subgroups are clearly unbroken and thus we can omit them in the following
analysis without affecting our conclusions. Then, under an infinitesimal
transformation on the vacuum we have
\begin{eqnarray}
\left({\displaystyle \sum_{i=1}^{8}}\alpha_{i}T_{i}+\gamma\, X\chi\textrm{1}_{3\times3}+\delta\,\textrm{PQ}_{\chi}\textrm{1}_{3\times3}\right)V_{\chi} & = & 0,\label{transformationVchi}\\
\left({\displaystyle \sum_{i=1}^{8}}\alpha_{i}T_{i}+\gamma\, X_{\rho}\textrm{1}_{3\times3}+\delta\,\textrm{PQ}_{\rho}\textrm{1}_{3\times3}\right)V_{\rho} & = & 0,\label{transformationVrho}
\end{eqnarray}
where $\alpha_{i}$, $\gamma$ and $\delta$ are independent real
constants and $1_{3\times3}$ denotes the $3\times3$ identity matrix.
Also, we have from Table \ref{U1charges} that $X\chi=-1/3$, $X_{\rho}=2/3$
and $\textrm{PQ}_{\chi}=\textrm{PQ}_{\rho}=1$. Since the $\chi$
and $\rho$ scalar triplets are in the fundamental representation
of $\textrm{SU(3)}_{L}$, the $T_{i}$ generators in Eqs.~(\ref{transformationVchi})
and (\ref{transformationVrho}) are given by $\lambda_{i}/2$, where
$\lambda_{i}$ are the well known Gell-Mann matrices. From Eqs.~(\ref{transformationVchi}-\ref{transformationVrho})
follows 
\begin{eqnarray}
v(\alpha_{1}-i\alpha_{2}) & = & 0,\label{eqforalpha1}\\
v(\alpha_{6}+i\alpha_{7}) & = & 0,\\
u\left(\alpha_{1}+i\alpha_{2}\right)+w\left(\alpha_{6}-i\alpha_{7}\right) & = & 0,\label{eqforalpha2}\\
v\left(-3\alpha_{3}+\sqrt{3}\alpha_{8}+4\gamma+6\delta\right) & = & 0,\label{eqforalpha4}\\
3u(\alpha_{4}+i\alpha_{5})-2w\left(\sqrt{3}\alpha_{8}+\gamma-3\delta\right) & = & 0,\label{eqforalpha5}\\
u\left(3\alpha_{3}+\sqrt{3}\alpha_{8}-2\gamma+6\delta\right)+3w(\alpha_{4}-i\alpha_{5}) & = & 0,\label{eqforalpha6}
\end{eqnarray}
with $i=\sqrt{-1}$. Solving simultaneously Eqs. (\ref{eqforalpha1}-\ref{eqforalpha6})
(with $u\neq0$, $v\neq0$ and $w\neq0$) we have that $\alpha_{1}=\alpha_{2}=\alpha_{5}=\alpha_{6}=\alpha_{7}=0$
and
\begin{eqnarray}
\alpha_{4} & = & -\frac{6uw}{u^{2}+w^{2}}\delta\equiv-3\sin(2\theta)\delta,\label{alpha4}\\
\alpha_{8} & = & \frac{6}{\sqrt{3}}\left(\frac{2w^{2}}{u^{2}+w^{2}}-1\right)\delta-\frac{\alpha_{3}}{\sqrt{3}}\equiv\frac{6\cos(2\theta)}{\sqrt{3}}\delta-\frac{\alpha_{3}}{\sqrt{3}},\\
\gamma & = & -\frac{3w^{2}}{u^{2}+w^{2}}\delta+\alpha_{3}\equiv-\frac{3}{2}\left(1+\cos(2\theta)\right)\delta+\alpha_{3},\label{delta}
\end{eqnarray}
where $\tan\theta\equiv u/w$. Since the parameters $\alpha_{3}$
and $\delta$ are independent, this implies that from the ten generators
only two linearly independent combinations, say $g_{1}$ and $g_{2}$,
remain unbroken. At a first glance, the choice of these generators
is arbitrary. However, we take into consideration that one of them
has to be the anomaly-free electric charge generator, which is achieved
by taken $\delta=0$ and $\alpha_{3}=1$. Doing so, $g_{1}=Q$. The
other generator, $g_{2}$, must have $\delta\neq0$ in order to be
linearly independent of $g_{1}$ (since all generator with $\delta=0$
will be proportional to $Q$). Hence, the unbroken generators are
written as:
\begin{eqnarray}
g_{1} & = & T_{3}-\frac{1}{\sqrt{3}}\, T_{8}+X,\\
g_{2} & = & \left[3\cos^{2}(\theta)\, T_{3}-3\sin(2\theta)\, T_{4}+\frac{1}{2}\sqrt{3}(3\cos(2\theta)-1)\, T_{8}+\textrm{PQ}\right]\mbox{\ensuremath{\delta}},\label{O2}
\end{eqnarray}
with $\delta\neq0$ in the last equation. The symmetry associated
to $g_{1}$, $\textrm{U(1)}_{Q}$, is anomaly free as it is well known.
The $g_{2}$ generator, which is independent from $g_{1}$, is a linear
combination of $T_{3},\, T_{4},\, T_{8},\:\mbox{PQ}$ generators.
We refer to the symmetry associated to $g_{2}$ as $\textrm{U(1)}_{H}$.
The key point here is that a part of the initial axial symmetry, $\textrm{U(1)}_{\textrm{PQ}}$,
remains unbroken because the coefficient $\delta$, in Eq. (\ref{O2}),
is always different from zero. In conclusion, the existence of $g_{1}$
and $g_{2}$ implies that the $\textrm{U(1)}_{Q}\otimes\textrm{U(1)}_{H}$
subgroup of the $\textrm{SU(3)}_{L}\otimes\textrm{U(1)}_{X}\otimes\textrm{U(1)}_{\textrm{PQ}}$
remains unbroken. 

Now, since $G$ is an exact symmetry, i.e. $\left[G,\,\mathcal{L}_{T}\right]=0$
(where $\mathcal{L}_{T}$ is the total Lagrangian of the model) from
the Goldstone's theorem \citep{Goldstone:1962es}, we have exactly
eight NG scalar bosons (a NG scalar boson for each broken generator),
which in this model will become the longitudinal component of the
eight massive gauge vector bosons via the Higgs mechanism. In the
physical scalar spectrum this model has only massive scalar bosons,
$H_{1}^{0},\, H_{2}^{0},\, H^{+},\, H^{-}$, as it is shown in \citep{Montero:2011tg}.
If the $g_{2}$ was broken, a NG scalar boson would appear in the
physical scalar spectrum. Because $g_{2}$ has a component in $\textrm{PQ}$
generator, that physical NG scalar boson would be an axion. However,
it does not happen and the model has three massless quarks (one $u-$type
quark and two $d-$type quarks) instead. This can be easily seen from
the mass matrices because a couple of rows in the $u-$quark mass
matrix and two couples of rows in the $d-$quark mass matrix are proportional
to each other, see Eqs. $\left(18\right)$ and $\left(19\right)$
in Ref. \citep{Montero:2011tg}. The exact form of those massless
quarks is neither clarifying nor relevant for our analysis, thus,
we do not write them here. These massless quarks are fully expected
because an exact axial symmetry that is realized in the WW manner
implies massless fermions. 

The action of the remaining $\textrm{U(1)}_{H}$ symmetry preventing
some quarks to gain mass becomes obvious when we change basis to work
with the mass eigenstates instead of the symmetry eigenstates. The
symmetry eigenstates $U_{iL,R},D_{jL,R}$ (with $i=1,\ldots,4$ and
$j=1,\ldots,5$) and the mass eigenstates $\left(U_{M}\right)_{iL,R},\left(D_{M}\right)_{jL,R}$
are related by $U_{L,R}=(V_{L,R}^{U})^{\dagger}\left(U_{M}\right)_{L,R}$
and $D_{L,R}=(V_{L,R}^{D})^{\dagger}\left(D_{M}\right)_{L,R}$, where
$V_{L,R}^{U,D}$ are independent unitary matrices such that $V_{L}^{U}M^{U}V_{R}^{U\dagger}=\hat{M}^{U}$
and $V_{L}^{D}M^{D}V_{R}^{D\dagger}=\hat{M}^{D}$, where $\hat{M}^{U}=\textrm{diag}(m_{u_{M1}},m_{u_{M2}},m_{u_{M3}},m_{u_{M4}})$
and $\hat{M}^{D}=\textrm{diag}(m_{d_{M1}},m_{d_{M2}},m_{d_{M3}},m_{d_{M4}},m_{d_{M5}})$.
Since the mass matrix of the $u-$ and $d$-quark types are not hermitian
matrices, in order to obtain the mass eigenstates we have to solve
the matrix equations: $V_{L}^{q}M^{q}M^{q\,\dagger}V_{L}^{q\dagger}=V_{R}^{q}M^{q\,\dagger}M^{q}V_{R}^{q\,\dagger}=(\hat{M}^{q})^{2},\, q=U,D.$
More specifically, we have to find the base-rotation matrices $V_{L,R}^{U}$
and $V_{L,R}^{D}$ to be able to write the Yukawa interactions in
terms of the quark-mass eigenstates. This task can be done by standard
procedures. Unfortunately, exact analytical expressions for these
matrix are enormous so that it is worthless to show them here. The
diagonalization study shows that we have one vanishing eigenvalue
in the $u-$quark sector and two in the $d-$quark sector, as expected.
The respective zero-mass eigenstates are clearly identified. Lets
call them as $u{}_{M1}$, $d{}_{M1}$, and $d{}_{M2}$. It means that
there are no mass terms of the form $m_{u{}_{M1}}\overline{{u{}_{M1}}_{L}}{u{}_{M1}}_{R}+\, m_{d{}_{M1}}\overline{{d{}_{M1}}_{L}}{d{}_{M1}}_{R}+\, m_{d{}_{M2}}\overline{{d{}_{M2}}_{L}}{d{}_{M2}}_{R}+\,\textrm{h.c.}$,
i.e. $m_{u{}_{M1}}=m_{d{}_{M1}}=m_{d{}_{M2}}=0$. 

We find more important results looking at the quark-scalar field interactions
coming from the Yukawa interactions in Eq. \ref{lagyuk}. Here we
find that there is no interactions involving the right component of
these massless quark states. The right states ${u{}_{M1}}_{R}$, ${d{}_{M1}}_{R}$,
and ${d{}_{M2}}_{R}$ disappear from the Yukawa interactions. In other
words, no left quark state are coupled with these massless states
through neutral- or charged-scalar fields. Nonetheless, these zero-mass
right components should have some interaction after all. In fact,
from the quark kinetic terms we find that they interact only with
the neutral vector bosons, $A_{\mu},\, Z_{\mu}$, and $Z_{\mu}^{\prime}$.
These interactions couple only quarks with the same chirality. It
means that each one of the right components ${u{}_{M1}}_{R}$, ${d{}_{M1}}_{R}$,
and ${d{}_{M2}}_{R}$ can be transformed by an arbitrary $\textrm{U}(1)_{H}$
phase without affecting any other term in the Lagrangian. Hence, looking
at the Yukawa and the neutral vector boson interactions, written in
terms of the quark-mass eigenstates, we can identify the $\textrm{U}(1)_{H}$
symmetry responsible for preventing these quark states to get mass:
the massless right fields ${u{}_{M1}}_{R}$, ${d{}_{M1}}_{R}$, and
${d{}_{M2}}_{R}$ transform as $e^{i\alpha}{u{}_{M1}}_{R},\, e^{i\alpha}{d{}_{M1}}_{R},\, e^{i\alpha}{d{}_{M2}}_{R}$
and all other fields transform trivially under $\textrm{U}(1)_{H}$
(note that this symmetry is anomalous with $\left[\textrm{SU(3)}_{C}\right]^{2}\textrm{U(1)}_{\textrm{H}}$
anomaly $\varpropto-3$). This is a clear and undoubtable manifestation
of the remaining symmetry we have found. If we consider perturbation
theory, these massless quarks can not get mass through radiative corrections
to their propagators since the right component of these fields disappeared
from the Yukawa interactions and they only couple to neutral vector
bosons which conserve chirality. Therefore, there is no way to form
loop diagrams to give mass for these particular fields, at all orders
in perturbation theory.

Now, let us discuss the possibility of generating masses for those
massless quarks through non-perturbative corrections and the viability
of this model to explain the low-energy hadron phenomenology. Roughly
speaking, that can be seen as follows. From both chiral QCD and lattice
calculations the ratio $\mu_{u}/\mu_{d}$ is $0.410\pm0.036$ \citep{Gasser:1982ap,Nelson:2003tb,Kim:2008hd},
where $\mu_{u}$ and $\mu_{d}$ are the \textquotedblleft low-energy
quark masses\textquotedblright . These should be distinguished from
the quark mass parameters, $m_{i}$ of the QCD Lagrangian at high
scale \citep{Choi:1988}. In particular, $\mu_{u}=\beta_{1}m_{u}+\beta_{2}\frac{m_{d}m_{s}}{\Lambda_{\chi SB}}$
where $\Lambda_{\chi SB}\sim1$ TeV (where we have identified naturally
the massless quarks as $u=u{}_{M1}$ and $d=d{}_{M1}$ and $s=d{}_{M2}$).
It means, $\mu_{\mu}$ receives an additive non-perturbative contribution
of order $m_{d}m_{s}$ in addition to the perturbative one, $\beta_{1}m_{u}$,
which is zero because both $m_{u}$ is zero. The non-perturbative
contribution is also zero because $m_{d}$ and $m_{s}$ are both zero
and $\beta_{2}$ is estimated to be a number of order one. Thus, $\mu_{u}=0$,
which is in complete disagreement with the ratio $\mu_{u}/\mu_{d}$.
A similar analysis is also valid for the $\mu_{d}$ and $\mu_{s}$
\citep{Banks1996}.

\section{Conclusions}

The scalar content in the E331 model is not enough to break down the
initial symmetry, $G$ to $\textrm{U(1)}_{Q}\otimes\textrm{U(1)}_{B}$.
Instead, an extra generator $g_{2}$ remains unbroken and thus the
model has a $\textrm{U(1)}_{H}$ axial symmetry after the spontaneous
symmetry breaking. As we have explicitly shown above, $g_{2}$ is
a linear combination of the $T_{3},\, T_{4},\, T_{8},\:\mbox{PQ}$
generators and it is linearly independent of the generators of the
electric charge and baryonic number, $g_{1}$ and $B$, respectively.
Because of the $\textrm{PQ}$ component in the $g_{2}$ generator,
we have that the initial axial $\textrm{U(1)}_{\textrm{PQ}}$ symmetry
is not completely broken. In other words the $\textrm{U(1)}_{Q}\otimes\textrm{U(1)}_{H}$
subgroup of the $\textrm{SU(3)}_{L}\otimes\textrm{U(1)}_{X}\otimes\textrm{U(1)}_{\textrm{PQ}}$
group remains unbroken. Therefore, the model has three massless quarks.
The $\textrm{U(1)}_{H}$ symmetry acts on the mass eigenstates as
an axial symmetry, $u{}_{M1R}\rightarrow e^{i\alpha}u{}_{M1R}$, $d{}_{M1R}\rightarrow e^{i\alpha}d{}_{M1R}$,
$d{}_{M2R}\rightarrow e^{i\alpha}d{}_{M2R}$, and this will protect
these massless quarks to acquire mass at any level of perturbation
theory. Furthermore, we recall that the unbroken $\textrm{U(1)}_{H}$
subgroup has its origin in an axial symmetry, $\textrm{U(1)}_{\textrm{PQ}}$,
which, although anomalous, is an accidental symmetry in the sense
that it follows from the gauge symmetries and renormalizability. Therefore,
the remaining axial symmetry acting on quarks will only be broken
down by non-perturbative QCD processes \citep{Hooft:1976up}. However,
these effects are not enough to provide the necessary low-energy quark
masses, $\mu_{i}$, to the three massless quarks to make the model
be in agreement with both chiral QCD and lattice calculations, which
give the ratio $\mu_{u}/\mu_{d}$ is $0.410\pm0.036$ \citep{Gasser:1982ap,Nelson:2003tb,Kim:2008hd}.
Hence, the economical version of the 3-3-1 model can not be considered
a realistic description of the electroweak interaction. 
\begin{acknowledgments}
B. L. S. V. would like to thank Coordenação de Aperfeiçoamento de
Pessoal de Nível Superior (CAPES), Brazil, for financial support under
contract 2264-13-7 and the Argonne National Laboratory for kind hospitality.
We are grateful to V. Pleitez for valuable discussions.
\end{acknowledgments}

\bibliographystyle{unsrt}
\bibliography{references}

\end{document}